\documentstyle[twoside,fleqn,npb,epsfig]{article}
\newcommand{\AmS}{{\protect\the\textfont2
  A\kern-.1667em\lower.5ex\hbox{M}\kern-.125emS}}

\title{The Minimal Left-Right Symmetric Model and Radiative Corrections to the Muon Decay}

\author{M. Czakon\address{Institut f\"ur Theoretische Physik, Universit\"at Karlsruhe, D-76128 Karlsruhe, Germany}, J. Gluza\address{DESY Zeuthen, Platanenallee 6, D-15738 Zeuthen, Germany}, 
J. Hejczyk\address{
    Institute of Physics, University of Silesia, Uniwersytecka 4, PL-40-007 Katowice,
    Poland}}  
\begin{document}

\begin{abstract}
A  self-consistent version of the left-right (LR) symmetric model is used to examine tree- as well as one-loop level
radiative corrections to the muon decay. It is shown that constraints on the heavy sector 
of the model parameters are different
when going beyond tree-level physics. In fact, in our case, the only useful constraints 
on the model can be obtained from the one-loop
level calculation. Furthermore, corrections coming from the subset of SM particles within the LR model have
a different structure from their SM equivalent, e.g. the top quark leading term contribution to $\Delta \rho$ within the LR model
is different from its SM counterpart. As a  consequence, care must be taken in  fitting procedures of models beyond the SM,
where usually, only tree-level couplings modified by the SM radiative corrections are considered. 
This procedure is not always correct.
\end{abstract}

\maketitle

\section{Introduction}
The smallest gauge group which implements the hypothesis of the left-right
symmetry of weak interactions is \cite{pati2}

\begin{equation}
\label{symmetry}
G_{LR}=SU(2)_L \otimes SU(2)_R \otimes U(1)_{B-L}.
\end{equation}

This gauge group can be understood as a second step (after the SM) 
in unifying fundamental interactions. The main feature of the model
is the restoration of both the quark-lepton  and parity symmetry.
At the same time the $U(1)$ generator  gets its physical interpretation as the 
B-L quantum number. Other phenomena which
are investigated are connected with small masses of light neutrinos,
charge quantization, understanding of  CP violation
in the quark sector, the strong CP problem, baryogenesis, etc.
Until present days literally hundreds of papers 
have been devoted to these concepts and their theoretical 
and phenomenological consequences.
An extended literature on the subject can be found
e.g. in the Introduction of \cite{ann}. 
The model is baroque
with many new particles of different types.
New neutral leptons,
charged and neutral gauge bosons, neutral and charged
Higgs particles appear.
There are many different versions of the LR models 
with the same or different left and right gauge couplings
$g_{L,R}$ and specific Higgs-sector representations. 
We chose the model with $g_L=g_R$ and a Higgs representation
with a bidoublet $\Phi$ and two (left and right) triplets $\Delta_{L,R}$.
We also assume that the VEV of the left-handed triplet $\Delta_{L}$ vanishes,
$<\Delta_{L}>=0$ and the CP symmetry is violated only by complex phases in quark
and lepton mixing matrices. We call this model the Minimal Left-Right Symmetric
Model (MLRM). Our aim is to show that constraints on the heavy sector of the model from 
muon decay at tree  and one loop levels are completely different.
First we will discuss  tree-level muon decay.
Bounds on $M_{W_2}$ (the additional charged gauge boson mass) from
this tree level process are cited permanently by PDG
\cite{pdg02}. We view 
the situation in the following way: 
a consistent model gives very weak  limits on charged 
current parameters from the tree level muon decay. 
As quite impressive  bounds derived from muon decay still persist 
through the succeeding PDG journals, we found it worth to clarify the case.
Then we go to the one-loop level results. We end up with conclusions and outlook.
 
\subsection{Muon decay at tree level: no bounds on charged current parameters}

As a low energy process, 
with a momentum transfer small relative to the involved gauge boson mass, the muon
decay can be conveniently described by a four-fermion interaction. 
For very small neutrino masses, neglecting the mixing between them,  the 
Lagrangian can be written in the form

\begin{eqnarray}
\label{1l}
-{\cal L} &=&\sum\limits_{i,j=L,R}\bar{c}_{ij} \bar{e} \gamma_{\alpha}P_i \nu_{e} 
   \bar{\nu}_{\mu}\gamma^{\alpha}P_j \mu . 
\end{eqnarray}
where ($g=g_L=g_R$):
\begin{eqnarray}
\bar{c}_{LL}&=&\frac{g^2}{2M_{W_1}^2} (\cos^2{\xi}+\beta \sin^2{\xi}),
 \\
\bar{c}_{RR}&=&\frac{g^2}{2M_{W_1}^2} (\sin^2{\xi}+\beta \cos^2{\xi}),
  \\
\bar{c}_{RL}=\bar{c}_{LR}&=&\frac{g^2}{2M_{W_1}^2} (-1+\beta)
\sin{\xi} \cos{\xi}. 
\end{eqnarray}
$\beta = \frac{M_{W_1}^2}{M_{W_2}^2}$,
$\xi$ is the mixing  between
the charged gauge bosons \cite{pati2,gluzzr1}.
Obviously, the $\beta \rightarrow 0$, $\xi \rightarrow 0$ limit leads to the SM result, with a purely
left-handed interaction.

To have neutrino mixings properly included, we have to write:
\begin{eqnarray}
-{\cal L} &=& \sum\limits_{i,j=L,R} \left( {c}_{ij} \right)_{ab} \bar{e} \gamma_{\alpha}P_i 
\nu_{a}   \bar{\nu}_{b}\gamma^{\alpha}P_j \mu  +{\cal L}_{heavy}, \nonumber
\end{eqnarray}
where:
\begin{eqnarray}
\left(  {c}_{ij} \right)_{ab} &=&\bar{c}_{ij}   (K_i^\dagger)_{e a} 
(K_j)_{b \mu},\;i,j=L,R. 
\end{eqnarray}

The matrices $K_{L,R}$ build up the neutrino mixing matrix $U$, which can be approximated
to be \cite{gluzzr1,hej}

\begin{equation}
U=  \left( \matrix{ K^T_L \cr K_R^{\dagger} } \right)=
 \left( \matrix{ O(1) & O(\frac{m_D}{M_N}) \cr
        O(\frac{m_D}{M_N}) & O(1)} \right).
\label{hlmix}
\end{equation}

$m_D$ is the Dirac neutrino mass matrix which emerges from vacuum expectation values (VEVs) in the  bidoublet 
Higgs-sector and $M_N$ stands for diagonal elements of the neutrino mass matrix $M_R$
connected with the right handed triplet Higgs representation  
(for details see e.g.  \cite{gluzzr1}).
The sum over $a$ and $b$ is understood, with both states light.
${\cal L}_{heavy}$ contains the sum over at least one heavy neutrino 
and for our purposes is irrelevant. 
We can see that apart from a pure left-handed term
$c_{LL}$, all others get extra damping factors connected with the $K_R$ mixing
matrix of light-heavy  neutrinos  being at most 
$\propto O(1\;{\rm GeV}/m_N) << 1$, 
 where $m_N$ is the lightest of heavy neutrinos ($m_N =  min(M_N)$). 
In what follows we consider  $m_N \geq 100$ GeV. 

In terms of the four-fermion interaction we can find 
\cite{sch,bos}:
\begin{equation}
8 G_F^2=|c_{LL}|^2+|c_{LR}|^2+|c_{RL}|^2+|c_{RR}|^2.
\end{equation}

Using  relations $c_{LL} \gg c_{RR},c_{LR},c_{RL}$
and $\sum\limits_{a=light} 
\left| {\left( K_L \right)}_{ea} \right|^2 \simeq 1$, 
we have:

\begin{eqnarray}
\frac{G_F}{\sqrt{2}} & \simeq & \frac{|c_{LL}|}{4} \\
&=& \frac{\pi \alpha}{2 s_W^2 M_{W_1}^2 
(1-\Delta r)}(cos^2{\xi}+\beta \sin^2{\xi}) \nonumber \\
& \simeq &
\frac{\pi \alpha}{2 s_W^2 M_{W_1}^2  
(1-\Delta r)}(1+\beta \xi^2). \nonumber
\end{eqnarray}

\begin{figure}
\epsfig{file=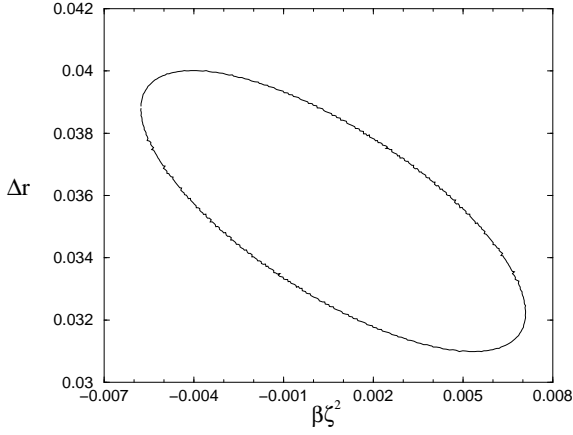, width=7.5cm}
\caption{90 \% C.L. region for the allowed $\beta \xi^2$ and $\Delta r$ parameters.}
\label{beta}
\end{figure}

To make the fitting procedure of the $\xi$ and $\beta$ parameters
possible at all at the tree-level,
we have to naively rely on SM corrections, we thus take  \cite{pdg02} 
$M_{W_1}=80.446\pm0.040$ GeV, $\Delta r=0.0355\pm0.0021$ and $s_W^2\equiv 
(s_W^2)_{SM}=1-\frac{M_{W_1}^2}{M_{Z_1}^2} = 0.2228 \pm 0.0004$. The result of the fit   
is plotted in Fig.~\ref{beta}.

Although $\beta\xi^2 < 0.007$ looks fine, 
with the most optimistic bound on $\xi$ below 0.1 \cite{pdg02,lowplb}, 
we get $\beta \leq 0.84$, i.e. $M_{W_2} \geq 1.2 M_{W_1} \simeq 100$ GeV. 

Let us finally note that if we only had light 
neutrinos  (Eq.~\ref{1l}) then much better bounds on  $M_{W_2}$ would be available
\cite{bos}.

Let us summarize. 
In a realistic LR model (i.e. when the mixing of
heavy Majorana neutrinos is taken into account), 
the tree-level diagrams for the muon decay give no
interesting bounds on $\beta$ (see also \cite{jap}).
Moreover, as it will be clear in the next Section, the procedure
we have used, where the SM values $\Delta r$ and $s_W^2$ 
have been taken into account, is wrong.

\begin{figure}[h]
\epsfig{file= 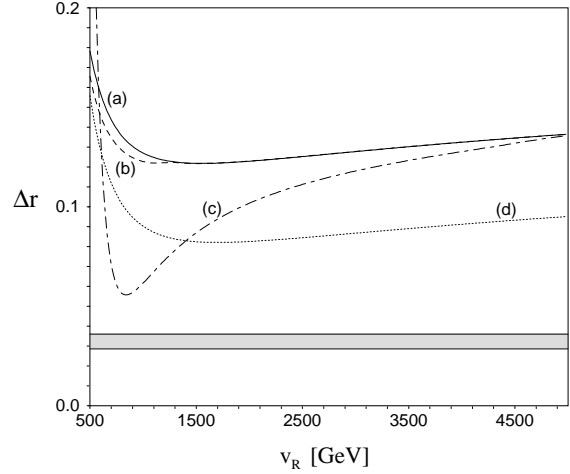, width=7.5cm}
\caption{
$\Delta r$ as function of $v_R$ for different heavy neutrino masses.
Higgs masses are chosen according to Eq.~\ref{ha}.  The (a)
line is for (three heavy neutrinos) $m_N=100$ GeV; (b) is for
$m_N=500$ GeV; (c)  is for $m_N=2$ TeV. Line (d) shows the results
when heavy neutrino masses follow from the maximal Yukawa coupling 
of the right-handed triplet Higgs-sector, $h_M=1$ \cite{hej}.
The gray area shows the  experimentally allowed values
of $\Delta r$ (SM prediction).}
\label{delta_r}
\end{figure}

\subsection{Constraints on the model parameters from the one-loop level}

Oblique radiative corrections to this process have been 
considered in the frame of the MLRM
in \cite{npb}. Further analysis has been given in \cite{hej}. 
Though the model has more free parameters (see e.g. \cite{ann,hej}), 
namely two gauge coplings $g=e/\sin{\Theta_W}$ 
and $g'=e/\sqrt{\cos{2\Theta_W}}$ altogether with three VEVs: 
$\kappa_1,\kappa_2$ (connected with the bidoublet $\Phi$) 
and $v_R$ (connected with 
the right handed triplet $\Delta_R$), 
there are simultaneously more physical
quantities ($e,M_{W_1},M_{W_2},M_{Z_1},M_{Z_2}$,) 
and unambiguous relations among them can be found 
($5 \to 5$ mapping).
This enables us to find (analogous to the SM)
the counterterm of
the sine squared of the Weinberg angle as function of
masses and their counterterms\footnote{For versions of the LR model
with more free parameters (e.g. $g_L \neq g_R$) the situation 
would be quite different: $s_W^2$ would not be predictable
in terms of gauge boson masses and their counter terms),
but would have to be tuned to experimental data).}
 \begin{eqnarray}
 \delta (s_W^2)_{LR} &  = & 2 c_W^2 
\frac{(\delta M_{Z_2}^2+\delta M_{Z_1}^2)}{\langle S \rangle } \nonumber \\
& - &  2 c_W^2 
\frac{(\delta M_{W_2}^2+\delta M_{W_1}^2)}
{\langle S \rangle } \nonumber \\ & & \nonumber \\
&+ & \frac{1}{2} \frac{(M_{W_2}^2+M_{W_1}^2)(\delta M_{Z_2}^2 
+\delta M_{Z_1}^2)}{\langle S \rangle ^2}   \nonumber \\
&+&\frac{1}{2} \frac{(M_{Z_2}^2+M_{Z_1}^2)(\delta M_{W_2}^2+\delta M_{W_1}^2) 
}{\langle S \rangle ^2} 
\nonumber \\
&- & \frac{1}{2} \frac{(2 M_{Z_1}^2+M_{Z_2}^2)\delta M_{Z_1}^2}{\langle S \rangle ^2} 
\nonumber \\ 
&-& \frac{1}{2} \frac{(2 M_{Z_2}^2+M_{Z_1}^2)\delta M_{Z_2}^2}
{\langle S \rangle ^2}. 
\end{eqnarray}

Let us note that the denominator $\langle S \rangle $ is proportional to the scale
of the right sector $v_R$
\begin{eqnarray}
\langle S \rangle  & \equiv & (M_{Z_2}^2+M_{Z_1}^2)-(M_{W_2}^2+M_{W_1}^2)
\nonumber \\
&  = & \frac{g^2}{2 \cos{2 \Theta_W} } v_R^2.
\end{eqnarray}
\\
$(\delta s_W^2)_{LR}$ exhibits a different structure from  the SM case.


\begin{figure}[h]
\epsfig{file= 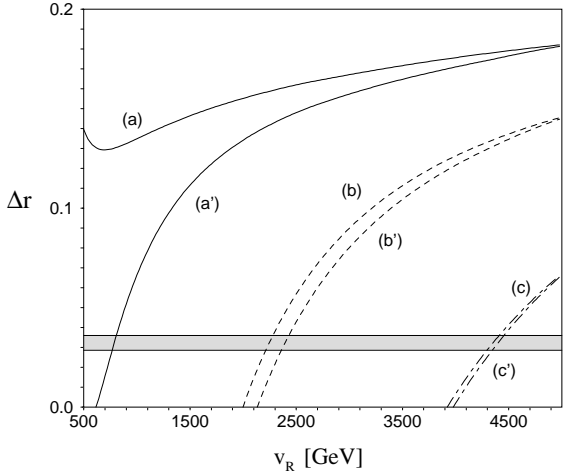, width=7.5cm}
\caption{$\Delta r$ as function of $v_R$. Sets with and without
primes show results for three heavy neutrino masses with $m_N=100$ GeV
and $m_N=2$ TeV respectively.  The lines describe different values of
Higgs scalar masses: (a) is for all Higgs masses  $M_{H}=1$ TeV; (b)
is for  $M_{H}=5$ TeV; (c) is for  $M_{H}=10$ TeV.}
\label{mn}
\end{figure}
\begin{figure}[h]
\epsfig{file= 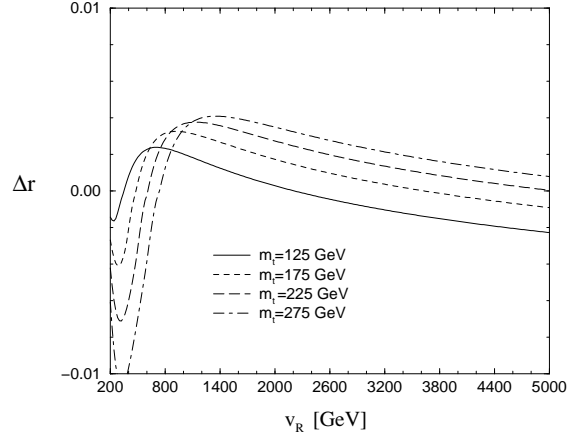, width=7.5cm}
\caption{The contribution of the third quark family to $\Delta r$ as 
function of $v_R$ for different top quark masses.}
\label{top}
\end{figure}

In  Figs.~\ref{delta_r}-\ref{top} the  contributions to the $\Delta r$ parameter defined as\footnote{To make possible a comparison to the SM result 
on $\Delta r$,  $\Delta r$
is modified to account for a different definition of the Weinberg
angle in both models \cite{hej} and  the relation 
$\frac{e^2}{(8 M_{W_1}^2 (s_W)^2_{SM})}(1+\Delta r) = \frac{e^2}{(8 M_{W_1}^2 (s_W)^2_{LR})}
(1+ \Delta r_{LR})$ is used. Let us add that not only $\delta s_W^2$ is different in 
LR and SM models, $\frac{\delta e}{e}$ has turned out to be a finite quantity \cite{npb,hej}.}

\begin{eqnarray}
 \Delta r &=& \frac{(s_W^2)_{SM}}{(s_W^2)_{LR}} (\Delta r)_{LR}+
\frac{(s_W^2)_{SM}}{(s_W^2)_{LR}}-1,  \nonumber \\
&& \nonumber \\
(\Delta r)_{LR}& =&  \left(
    \frac{-\Pi^T_{W_1}(0)-\delta M_{W_1}^2}{M_{W_1}^2}
    +2\frac{\delta e}{e} \right. \nonumber \\
&-& \left.  \frac{(\delta s_W^2)_{LR}}{(s_W^2)_{LR}}+\delta  \right) 
\end{eqnarray}

are  given. $\delta$  denotes the complete  vertex, box 
and external line corrections within the MLRM. 

If we parametrize the Higgs scalar masses by (no fine-tuning in the Higgs potential \cite{hej,gun}) 
\begin{eqnarray}
  M_{H_a} & \equiv & M_{H_1^0}=M_{H_3^0}=M_{A_1^0}=M_{A_2^0} \label{ha} \\ 
  &=&M_{H_1^+}=M_{H_2^+}=M_{\delta_L^{++}} = v_R/\sqrt{2},  \nonumber \\
  M_{H_b} & \equiv  & M_{H_2^0}=M_{\delta_R^{++}}=\sqrt{2} v_R,   \;\;
  M_{H_0^0} = \sqrt{2} \kappa_1 \nonumber 
\end{eqnarray}

then we can observe from Fig.~\ref{delta_r} that the experimental data on the
muon decay lifetime can not be accomodated. It is possible, however,  if 
all heavy Higgs particle masses are equal (see Fig.~\ref{mn}). 
Line (d) in Fig.~\ref{delta_r} shows the results
when heavy neutrino masses follow from the maximal Yukawa coupling 
connected with right-handed triplet representation $h_M=1$,
$m_N = \sqrt{2} v_R$ \cite{hej}.
 
For $h_M > 1$ the perturbative theory  breaks, which can be seen if the 
box diagrams are 
considered  \cite{hej}. In the model under investigation the 
light-heavy neutrino mixing has been neglected and the light-heavy gauge boson 
mixing angle $\xi$ is neglected. 
These assumptions are well motivated phenomenologically \cite{hej,app}. 
Fig.~\ref{top} shows explicitly that $\Delta r$  
strongly depends on the relation between  $m_t$ and $v_R$.
This means that  $m_t$  can not be predicted in the MLRM model 
without knowledge of the  
$v_R$ scale and furthermore that for larger $v_R$ the dependence will lead only to a very
crude bound.

The results shown here (for details, see \cite{hej,npb}) justify again 
our statements considered in \cite{epj}.
It has been concluded there, that the only sensible way to confront a model beyond the SM with the experimental data is 
to renormalize it self-consistently as it does not necessarily embed
 the SM structure of radiative corrections. If this is
not done, parameters which depend strongly on quantum effects should be left free in fits, though essential physics is
lost in this way. 

\section{Conclusions}

In LR models there are several new extra parameters (e.g. mixing
angles in the gauge sector, the $g'$ gauge coupling) along with quite
a lot of new particles and interactions. These cause that the model is a very 
good theoretical lab for examining many phenomenological
problems and issues of fundamental interactions. However, the freedom
of parameter space connected with the extra sector is not
unlimited, moreover, sometimes the model can be even more 
restricted than the SM alone. This seems to be particularly
true when processes are considered at the loop level.
Though we have restricted ourselves to the case of the minimal LR model,
the results already show that fine-tuning of the heavy sector parameters
must be done to recover experimental data.
This is in our opinion the main direction of future investigations
which has certainly not been fully exploited in the past \cite{oth}. 

\section*{Acknowledgements}

M. C. would like to thank the Alexander von Humboldt foundation for
fellowship.  This work was partly supported by the Polish Committee
for Scientific Research under Grant No. 2P03B05418 and by the European 
Community's Human Potential Programme under contract HPRN-CT-2000-00149
Physics at Colliders.

\newpage

\end{document}